\documentclass{llncs}
\usepackage{amssymb}

\newtheorem{openquestion}{Open Question}
\title{Large Isolating Cuts Shrink the Multiway Cut}     
\author{Igor Razgon\\ {\small ir45@mcs.le.ac.uk}}
\institute{Department of Computer Science, University of Leicester}
\begin{document}
\maketitle
\begin{abstract}
We propose a preprocessing algorithm for the multiway cut problem 
that establishes its polynomial kernelizability when the difference 
between the parameter $k$ and the size of the smallest isolating 
cut is at most $log(k)$.
To the best of our knowledge, this is the first progress
towards kernelization of the multiway cut problem. 
We pose two open questions that, if answered
affirmatively, would imply, combined with the proposed result,
unconditional polynomial kernelizability of the multiway cut problem.
\end{abstract}

\section{Introduction}
{\bf 1.1. Overview of the proposed results.}
Given a pair $(G,T)$ where $G$ is a graph 
and $T$ is a specified set of vertices, called \emph{terminals},
a (vertex) Multiway Cut (\textsc{mwc}) of $(G,T)$ is a set
of non-terminal vertices whose removal from $G$ separates all
the terminals.  The \textsc{mwc} problem asks to compute the
smallest \textsc{mwc} of $G$. It is NP-hard for $|T| \geq 3$ \cite{Dahlapprox}.

In this paper we concentrate on the parameterized version $(G,T,k)$ of
the \textsc{mwc} problem where we are given a parameter $k$ and asked
whether there is an \textsc{mwc} of $(G,T)$ of size at most $k$.
The goal we work towards is understanding the \emph{kernelizability}
of the \textsc{mwc} problem. In other words, we want to understand,
whether there is a polynomial time algorithm that transforms
$(G,T,k)$ into an \emph{equivalent} instance $(G',T',k')$ (equivalent in
the sense that the former is the 'YES' instance iff the latter is)
such that $|V(G')|$ is upper-bounded by a polynomial of $k'$
and $k'$ itself is upper bounded by a polynomial of $k$. Informally speaking
we want to shrink the instance of the \textsc{mwc} problem to a size
polynomially dependent on the parameter.

The kernelizability of the \textsc{mwc} is considered by
the parameterized complexity community as an interesting and challenging
question. In this paper we propose a partial result and pose two open
questions that, if resolved affirmatively, will imply, together with
this result, that the \textsc{mwc} problem is kernelizable. An informal
overview is given below.

Let $(G,T,k)$ be an instance of the \textsc{mwc} problem. 
An \emph{isolating cut} \cite{Dahlapprox} is a set of non-terminal 
vertices separating a terminal $t$ from the rest of terminals. 
Let $r$ be the smallest size of an isolating cut. Clearly we can assume 
$r \leq k$ otherwise, $(G,T,k)$ is a 'NO' instance. In this paper we propose
an algorithm transforming the initial instance into an equivalent one 
whose size is $O(2^{k-r}rk^2)$. The runtime of this algorithm is
$O(A(n)+2^{k-r}n^3r^2k^4)$ where $A(n)$ is the runtime of the constant ratio
approximation algorithm for the vertex \textsc{mwc} problem proposed in 
\cite{Gargapprox}. Thus we demonstrate that for every fixed constant
$c$ the subclass of \textsc{mwc} problem consisting of  instances with 
$k-r \leq c*logk$ is polynomially kernelizable. 
To the best of our knowledge, this is the first progress towards kernelization
of the \textsc{mwc} problem. 
Two more merits of the
proposed results are that it might be a building block in an unconditional
kernelization of the \textsc{mwc} problem and that it gives a new insight
into the structure of \emph{important separators} \cite{MarxTCS}.
To justify these merits, we provide below a more detailed overview of the
proposed result.

The main ingredient of the proposed algorithm is computing for each
$t \in T$ the union $U_t$ of all important isolating cuts of $t$ of size
at most $k$. An almost immediate consequence of Lemma 3.6. of \cite{MarxTCS}
shows that the union of all $U_t$ contains a solution of $(G,T,k)$ if
such exists. Therefore, 'contracting' the rest of non-terminal vertices
results in an instance equivalent to $(G,T,k)$. Prior to computing
the sets $U_t$ we ensure that the size of $|T|$ is at most $2k(k+1)$.
This is done in Section 3 by running the approximation algorithm 
of \cite{Gargapprox}
and processing the output in the flavour of a simple quadratic kernelization
algorithm for the Vertex Cover problem (i.e. noticing that the vertices
of the given \textsc{mwc} adjacent to a large number of terminal
components must be present in any solution and, 
after removal of these vertices and the already separated
terminals, the number of remaining terminals is small).

But what is the size of $U_t$ and what is the time needed for its computation?
To understand this, we study (in Section 2) 
important $X-Y$ separators of $G$ \cite{MarxTCS} where $X$ and $Y$
are two arbitrary subsets of vertices. As a result, we obtain a combinatorial
theorem saying that if $r$ is the smallest size of an $X-Y$ separator
then for an arbitrary $x$ the size of the union of all $X-Y$ important
separators of size at most $r+x$ is at most $2^{x+1}r$ and these vertices
can be computed in time $O(n^32^xr^2(r+x)^2)$. \footnote{The results are obtained
without any regard to the \textsc{mwc} problem, hence they might be of an
independent interest.} The exponential
part of the runtime follows from the need to enumerate so-called
\emph{principal important separators} whose union includes
all the needed vertices. We argue that the principal important separators
constitute a generally small subset of the whole set of important
separators and pose the \emph{first open question} asking whether
the number of principal separators can be bounded by a polynomial of $n$.
The affirmative answer to this question implies the polynomial runtime
of the algorithm proposed in this paper. In this case the algorithm can
be a first step of a kernelization method of the \textsc{mwc}.
However, it cannot be \emph{the only} step. We demonstrate the 
upper bound on the number of vertices is tight and hence generally
cannot polynomially depend on $k$. Therefore, a natural question is
whether the output of this algorithm can be further processed to obtain
an unconditional kernelization. We pose this as our \emph{second open
question}.

{\bf 1.2. Related work.} There are many publications related to the 
topics considered in the paper. We overview only those that are of 
a direct relevance for the proposed results.

The fixed-parameter tractability of the \textsc{mwc} problem has been
established in \cite{MarxTCS} and the runtime has been improved
to $O^*(4^k)$ in \cite{ChenAlgorithmica}. A parameterization
of the \textsc{mwc} problem above a guaranteed value has been recently
proposed in \cite{REX}, where we show that the problem is in XP under
this parameterization leaving open the fixed-parameter tractability status.

The notion of \emph{important separator} has been introduced in 
\cite{MarxTCS}. As noticed in \cite{LokshtanovClustering}, 
the recent algorithms for a number of challenging graph separation
problems, including the one of \cite{ChenAlgorithmica}, are
based on enumeration of important separators. Further on,
\cite{LokshtanovClustering} proves an upper bound $4^k$
on the number of important separators of size at most $k$ and 
notices that the algorithm of  \cite{ChenAlgorithmica} 
in fact implicitly establishes this upper bound. 
An alternative upper bound, suitable for the case where the 
smallest important separator is large, is established in \cite{REX}.

Constant ratio approximation algorithms for the \textsc{mwc}
problem have been first proposed in \cite{Dahlapprox} for
the edge version and in \cite{Gargapprox} for the vertex
version.

The research on kernelization has been given its current
shape by the landmark paper \cite{Bodnokernel}, which
allowed to classify fixed-parameter tractable problems
into kernelizable ones and those that are probably not. Among the many known 
kernelizability and non kernelizability results, 
let us mention the kernelization methods for multicut for trees 
\cite{DaligaultSTACS09} and Feedback Vertex Set \cite{FVSkernel} and non-kernelizability
proof for the Disjoint Cycles problem \cite{Bod2nokernel}. 
Although far from being analogous to the \textsc{mwc} problem,
all these problems are related to the flow maximization/cut minimization
tasks and hence might be a source of ideas useful for the final 
settling of the kernelizability of \textsc{mwc} problem.

\section{Bounding the union of important separators}
Let $X$ and $Y$ be two disjoint sets of vertices of the given graph $G$.
A set $K \subseteq V(G) \setminus (X \cup Y)$ is an $X-Y$ separator
if in $G \setminus K$ there is no path from $X$ to $Y$.
Let $A,B$ be two disjoint subsets of $V(G)$. We denote
by $NR(G,A,B)$ the set of vertices that are not reachable from $A$ in $G \setminus B$
Let $K_1$ and $K_2$ be two $X-Y$ separators. We say that $K_1 \prec^* K_2$ if
$NR(G,Y,K_1) \subset NR(G,Y,K_2)$.

A minimal $X-Y$ separator $K$ is called \emph{important} if there is no $X-Y$ separator $K'$
such that $K \prec^* K'$ and $|K| \geq |K'|$. This notion was first introduced in \cite{MarxTCS}
in a slightly different although equivalent way (see Proposition 3 of \cite{REX}). 
Let $r$ be the size of a smallest important $X-Y$ separator and let $S$ be an arbitrary
important separator. We call $|S|-r$ the \emph{excess} of $S$. Then the following theorem
holds.

\begin{theorem} \label{MainResult}
Let $U$ be the union of important $X-Y$ separators of excess at most $x$.
Then $|U| \leq 2^{x+1}r$. Moreover, $U$ can be computed in time
$O(n^32^{2x}r^2(r+x)^2)$.
\end{theorem}

In this section we prove Theorem \ref{MainResult} and show the tightness
of the upper bound of $|U|$.
The proof of  Theorem \ref{MainResult} in divided into two stages.
On the first stage we introduce a partially ordered family of subsets of the given
set satisfying a number of certain properties. 
We call such family of sets an \emph{IS-family}.
We prove Theorem \ref{MainResult} in terms of the IS family. Then we
show that the family of all important separators with the $\prec^*$ relation 
is in fact an IS family from where  Theorem \ref{MainResult} immediately follows.

The advantage of such 'axiomatic' way of proof is the possibility to clearly
specify the properties of the family of important separators (viewed as a 
partially ordered family of sets) that imply the above upper bound. 
An additional potential advantage is that some deep algebraic techniques
might become applicable for further investigation of the kernelization of
multiway cut. 

\subsection{From Important Separators to Partially Ordered Families of Sets}
Let $V$ be a finite set.
Let $({\bf F}, \prec)$ be a pair where ${\bf F}$ is a family
of subsets of $V$ and $\prec$ is an order relation on
the elements of ${\bf F}$ . Let $S \in {\bf F}$ and $v \in V$.
We say that $S$ \emph{covers} $v$ if
there is $S' \in {\bf F}$ such that
$S' \prec S$ and $v \in S' \setminus S$.
We define $Pred(S)$ to be the set of
all $S'$ such that $S' \prec S$ and
there is no $S'' \in {\bf F}$ such that
$S' \prec S'' \prec S$. Symmetrically,
we define $Succ(S)$ to be the set of
all $S' \in {\bf F}$ such that $S \prec S'$
and there is no $S'' \in {\bf F}$ such that
$S \prec S'' \prec S'$.
We define the \emph{visible set} of $S$ denoted
by $Vis(S)$ to be the set of all $v \in V$ satisfying
the following two conditions:
\begin{itemize}
\item there is $S' \in Pred(S)$ such that $v \in S'$;
\item $v$ is not covered by any element of $Pred(S)$.
\end{itemize}

$({\bf F}, \prec)$ is called an $IS$-family
if the following conditions are true.

\begin{itemize}
\item {\bf Smallest element (SE) condition.}
There is a unique element of ${\bf F}$ denoted
by $sm({\bf F})$ such that for any other
$S \in {\bf F}$, $sm({\bf F}) \prec S$.
\item {\bf Strict monotonicity (SM) condition.}
Let $S_1,S_2 \in {\bf F}$. If $S_1 \prec S_2$
then $|S_1|<|S_2|$. 
\item {\bf Single witness (SW) condition.}
Let $S \in {\bf F}$ and let $v \in S$.
Let $S'$ be a minimal element such that
$S \prec S'$ and $v \in S \setminus S'$.
We call $S'$ a \emph{witness} of $v$ w.r.t.
$S$. The condition requires that there is 
\emph{at most one} witness of $v$ w.r.t. $S$.
\item {\bf Transitive Elimination (TE) condition}
Let $S_1 \prec S_2 \prec S_3$ be three elements
of ${\bf F}$ and let $v \in S_1 \setminus S_2$.
Then $v \in S_1 \setminus S_3$.
\item {\bf Large visible set (LVS) condition}
Let $S \in {\bf F}$ and let $S' \in Pred(S)$.
Then $|S'| \leq |Vis(S)|$. For the subsequent
proofs we will use the \emph{extended} {\bf LVS}
condition stating that for each $S'' \prec S$,
$|S''| \leq |Vis(S)|$, which immediately follows
from the combination of {\bf LVS} and {\bf SM}
conditions.
\item {\bf Distinct visible set (DVS) condition}
For each $S \in {\bf F}$ such that $S \neq sm({\bf F})$.
Then $Vis(S) \not\subset S$.
\item {\bf Efficient Computability (EC) condition}
Let $n=|V|$. In $O(n^3)$ we can compute $sm({\bf F})$ 
as well as the witness of $v$ w.r.t. $S$ for the
given $S \in {\bf F}$ and $v \in S$ (or return 'NO' in case such
witness does not exist). The relation $S_1 \prec S_2$
can be tested in $O(|S_1|)$.
\end{itemize}

In the rest of this subsection we assume that $({\bf F}, \prec)$
is an IS-family. Our reasoning consists of three stages.
On the first stage, we prove 3 propositions stating simple properties
of an IS family. On the second stage we prove Theorem \ref{MainLevelBound},
our main counting result. The main body of the proof is provided in the
3 preceding lemmas. On the last stage we prove an analogue of Theorem \ref{MainResult}
for IS families: Corollary \ref{VertexBound} proves the upper bound on the
size of the union of the respective sets and Theorem \ref{runtime} establishes
an algorithm for computing of these sets. 

For $S \in {\bf F}$ let us denote $S \setminus \bigcup_{S' \prec S} S'$ by $hat(S)$.

\begin{proposition} \label{PropHat}
$hat(S)=S \setminus Vis(S)$.
\end{proposition}

{\bf Proof.}
It is clear from the definition that $hat(S) \subseteq S \setminus Vis(S)$.
Conversely, consider $v \in (S \setminus Vis(S)) \setminus hat(S)$. What can we say
about such $v$? First, that $v \in S$. Then, since $v \in \bigcup_{S' \prec S} S' \setminus Vis(S)$,
there may be two possibilities. According to one of them,
$v \in S' \prec S$ such that $S' \notin Pred(S)$ and $v$ does 
not belong to any $S'' \in Pred(S)$. It follows that there is $S'' \in Pred(S)$ such that
$S' \prec S''$ and $v \in S' \setminus S''$. Since $S'' \prec S$, $v \notin S$ by the 
{\bf TE} condition in contradiction to our assumption. The other possibility may be that
$v \in S' \in Pred(S)$ and $v$ is covered by another $S'' \in Pred(S)$. Then analogous reasoning applies.
By definition of a covered vertex, there is $S^* \prec S''$ such that $v \in S^* \setminus S''$ and again
$v \notin S$ by the {\bf TE}  condition, yielding an analogous contradiction. $\blacksquare$.

\begin{proposition} \label{WitnessBasic}
Let $S_1,S_2,v$ be such that $S_1 \prec S_2$ and $v \in S_1 \setminus S_2$.
Let $S^*$ be the witness of $v$ w.r.t. $S_1$. Then $S^* \preceq S_2$.
\end{proposition}

{\bf Proof.}
Let $S''$ be a minimal element of ${\bf F}$ such that $S_1 \prec S'' \preceq S_2$
and $v \in S_1 \setminus S''$. Then $S''$ is a witness of $v$ w.r.t. $S_1$.
By the {\bf SW} condition, $S''=S^*$. $\blacksquare$

\begin{proposition} \label{WitnessAdvanced}
Let $S \in {\bf F}$ and let $v \in Vis(S) \setminus S$. 
Then there is $S^* \prec S$ such that $v \in hat(S^*)$ and 
$S$ is the witness of $v$ w.r.t. $S^*$.
\end{proposition}

{\bf Proof.}
Let $S^*$ be a minimal element of ${\bf F}$ preceding $S$ such that $v \in S^*$.
Then $v \in hat(S^*)$. Indeed, otherwise, there is $S'$ such that 
$v \in S' \prec S^* \prec S$ in contradiction
to the choice of $S^*$. Assume by contradiction that $S$ is 
not the witness of $v$ w.r.t. $S^*$ and let $S''$ be this witness. 
According to Proposition \ref{WitnessBasic}, $S'' \prec S$. Let $S_2 \in Pred(S)$ be such that
$S'' \preceq S_2$. Clearly $S^* \prec S_2$. If $S''=S_2$ then $v \in S^* \setminus S_2$
by definition of $S''$. Otherwise, $v \in S^* \setminus S_2$ by the {\bf TE}
condition. It follows that $S_2$ covers $v$. Consequently, $v \notin Vis(S)$, a contradiction
proving that $S$ is indeed the witness of $v$ w.r.t. $S^*$. $\blacksquare$

For $S \in {\bf F}$, let's call $|S|-|sm({\bf F})|$, the \emph{excess} of $S$
and denote it $ex(S)$. 

\begin{lemma} \label{BasicBound}
Let $S \in {\bf F}$ such that $S \neq sm({\bf F})$.
For $v \in Vis(S) \setminus S$,let $S(v)$ be such
that $v \in hat(S(v))$ and $S$ is the witness of $v$ w.r.t. $S(v)$
(the existence of such $S(v)$ follows from Proposition \ref{WitnessAdvanced}).
Then $|hat(S)| \leq \sum_{v \in Vis(S) \setminus S} 2^{ex(S)-ex(S(v))}$.
\end{lemma}

{\bf Proof.}
If $Vis(S)=S$ then by Proposition \ref{PropHat}, $hat(S)=\emptyset$ and
we are done. Otherwise, the {\bf DVS} condition
allows us to fix a $v^* \in Vis(S) \setminus S$ .
Let us define a function $f$ on $V$ as follows:
$f(v^*)=2^{ex(S)-ex(S(v^*))}$ and for $w \neq v^*$, $f(w)=1$.
For $S \subseteq V$, the function naturally extends to 
$f(S)=\sum_{v \in S} f(v)$.

\begin{claim}
$|hat(S)| \leq f(Vis(S) \setminus S)$
\end{claim}
Observe that
$f(Vis(S))=|Vis(S)\setminus \{v^*\}|+f(v^*)=|Vis(S)|+f(v^*)-1$.
By the extended {\bf LVS} condition, the rightmost part 
of the above equality does not increase if we replace 
$Vis(S)$ by $S(v)$, i.e. $f(Vis(S)) \geq |S(v)|+f(v^*)-1$. 
Since $ex(S)-ex(S(v)) \geq 1$, by the {\bf SM}  condition,
$f(v^*) \geq ex(S)-ex(S(v))+1$. That is, $f(Vis(S)) \geq 
|S(v)|+ex(S)-ex(S(v))=|S|$.
Furthermore $f(Vis(S))=f(Vis(S)\setminus S)+f(Vis(S) \cap S)=f(Vis(S) \setminus S)+|Vis(S) \cap S|$.
On the other hand, $|S|=|S \setminus Vis(S)|+|Vis(S) \cap S|=|hat(S)|+|Vis(S) \cap S|$, the last
equality follows from Proposition \ref{PropHat}.
Thus the desired claim follows by removal $|Vis(S) \cap S|$ from the both sides of the 
inequality $f(Vis(S)) \geq S$. $\square$ 

Observe that due to the {\bf SM}  condition, for each $v \in Vis(S) \setminus S$,
$2^{ex(S)-ex(S(v))} \geq f(v)$, hence $f(Vis(S) \setminus S) \leq \sum_{v \in Vis(S) \setminus S} 2^{ex(S)-ex(S(v))}$. 
Therefore the lemma  follows from the above claim.
$\blacksquare$

For $x \geq 0$, let ${\bf E}_x$ be the subset of ${\bf F}$ consisting
of all the elements of excess at most $x$.
Let $S \in {\bf E}_x$. The $x$-hat of $S$ denoted by $hat_x(S)$
is a subset of $hat(S)$ consisting of all elements $v$ such that
there is no $S' \in {\bf E}_x$ such that $S \prec S'$ and $v \in S \setminus S'$.

\begin{lemma} \label{FirstLevelBound}
For any $x \geq 0$ \\
$\sum_{S \in {\bf E}_x} 2^{x-ex(S)+1}*|hat_x(S) \setminus hat_{x+1}(S)| 
\geq \sum_{S' \in {\bf E}_{x+1} \setminus {\bf E}_x} |hat(S')|$.
\end{lemma}

{\bf Proof.}
Denote the elements of ${\bf E}_x$ by $S_1, \dots, S_m$.
Denote $\{(v,i)| 1 \leq i \leq m, v \in hat_x(S_i) \setminus hat_{x+1}(S_i)\}$
by $OS$. For each $(v,i) \in OS$, let $aw(v,i)=2^{x-ex(S_i)+1}$.
For $OS' \subseteq OS$, let 
$aw(OS')=\sum_{(v,i) \in OS'} aw(v,i)$. It is not hard to
see that the left part of the desired inequality is 
$aw(OS)$. Indeed, for the given $i$, 
if we sum up $aw(v,i)$ for all $(v,i) \in OS$ 
then the total amount will be exactly
$2^{x-ex(S_i)+1}*|hat_x(S)|$.

Consider $(v,i) \in OS$. 
Then, since $v \notin hat_{x+1}(S_i)$, there is $S' \in {\bf E}_{x+1} \setminus {\bf E}_x$
such that $S_i \prec S'$ and $v \in S_i \setminus S'$.
We claim that $S'$ is in fact the witness of $v$ w.r.t. $S_i$.
Indeed, otherwise, according to Proposition \ref{WitnessBasic}, $S'$ succeeds
the witness of $v$ w.r.t. $v$ hence, by the {\bf SM}  condition,
the size of the latter is at most $x$. However, this contradicts $v \in hat_x(S_i)$.
By the {\bf SW} condition, the above $S'$ is unique for $(v,i)$.
So, we can say that $(v,i)$ is \emph{witnessed} by $S'$.

Denote the elements of ${\bf E}_{x+1} \setminus {\bf E}_x$ by $S'_1, \dots, S'_q$.
Partition $OS$ into $OS_1, \dots, OS_q$ such that the elements of $OS_q$ are
witnessed by $S'_q$. To confirm the lemma, it remains to prove that, for the given $i$,
$aw(OS_i) \geq |hat(S'_i)|$. 

Let $v \in Vis(S'_i)$. According to Proposition \ref{WitnessAdvanced},
there is $S^* \prec S'_i$ such that $v \in hat(S^*)$ and $S'_i$
is the witness of $v$ w.r.t. $S^*$. By the {\bf SM}
condition $ex(S^*) \leq x$, that is $S^* \in {\bf E}_x$. 
Observe that in fact $v \in hat_x(S^*) \setminus hat_{x+1}(S^*)$.
Indeed, otherwise there is an element $S''$ of ${\bf E}_x$
such that $S^* \prec S''$ and $v \in S^* \setminus S''$. But then
$S \prec S''$ by Proposition \ref{WitnessBasic} in contradiction to the 
{\bf SM} condition. Let $j(v)$ be such that $S^*=S_{j(v)}$.
It follows that $(v,j(v)) \in OS_i$. Consequently, 
$aw(OS_i) \geq \sum_{v \in Vis(S'_i)} 2^{x+1-ex(S_{j(v)})} \geq 
\sum_{v \in Vis(S'_i)} 2^{ex(S'_i)-ex(S_{j(v)})} \geq |hat(S'_i)|$,
the last inequality follows from Lemma \ref{BasicBound}. 
$\blacksquare$

For $x \geq 0$, denote $\sum_{S \in {\bf E}_x} 2^{x-ex(S)}|hat_x(S)|$
by $M(x)$. Then the following statement takes
place.

\begin{lemma} \label{SecondLevelBound}
For each $x \geq 0$, $M(x+1) \leq 2M(x)$.
\end{lemma}

{\bf Proof.}
First of all, observe that for each $S \in {\bf E}_{x+1} \setminus {\bf E}_x$,
$hat_{x+1}(S)=hat(S)$ just because, by the {\bf SM}  condition, 
there is no $S' \in {\bf E}_{x+1}$ such that $S \prec S'$. Furthermore, by definition,
the excess of $S$ is $x+1$. Therefore $|hat(S)|=2^{x+1-ex(S)}|hat_{x+1}(S)|$.
That is, we can rewrite the inequality of 
Lemma \ref{FirstLevelBound} as\\
$\sum_{S \in {\bf E}_x} 2^{x-ex(S)+1}*
|hat_x(S) \setminus hat_{x+1}(S)| \geq 
 \sum_{S' \in {\bf E}_{x+1} \setminus {\bf E}_x} 2^{x+1-ex(S)}|hat_{x+1}(S')|$

Furthermore, observe that for each $S \in {\bf E}_x$, $hat_{x+1}(S) \subseteq hat_x(S)$,
therefore $hat_{x+1}(S)=hat_{x+1}(S) \cap hat_x(S)$. Then we can safely add 
$\sum_{S \in {\bf E}_x} 2^{x-ex(S)+1}* |hat_x(S) \cap hat_{x+1}(S)|$ to the
left part of the inequality of the previous paragraph and 
$\sum_{S \in {\bf E}_x} 2^{x-ex(S)+1}*|hat_{x+1}(S)|$ 
to the right part of this inequality. Then after noticing that for each $S \in {\bf E}_x$,
$|hat_x(S) \cap hat_{x+1}(S)|+|hat_x(S) \setminus hat_{x+1}(S)|=|hat_x(S)|$ and that the
right part in fact explores $|hat_{x+1}(S')|$ for all elements $S' \in {\bf E}_{x+1}$,
the resulting inequality is transformed into:\\
$\sum_{S \in {\bf E}_x} 2^{x-ex(S)+1}* |hat_x(S)| \geq
 \sum_{S' \in {\bf E}_{x+1}} 2^{x-ex(S')+1}*|hat_{x+1}(S')|$.
It remains to notice that the left part of this inequality is $2M(x)$
and the right part is $M(x+1)$. $\blacksquare$

Now we are ready to state the main counting result.

\begin{theorem} \label{MainLevelBound}
For each $x \geq 0$, $M(x) \leq 2^x|sm({\bf F})|$.
\end{theorem}

{\bf Proof.}
Applying inductively Lemma \ref{SecondLevelBound}, it is easy to see
that $M(x) \leq 2^xM(0)$. By definition,
$M(0)=\sum_{S \in {\bf E}_0} 2^{0-ex(S)} |hat_0(S)|$. Since the only
element of ${\bf E}_0$ is $sm({\bf F})$ whose excess is $0$ and
$hat_0(sm({\bf F}))=sm({\bf F})$, the theorem follows. $\blacksquare$

The following corollary is the first statement of
Theorem \ref{MainResult} in terms of an IS family

\begin{corollary} \label{VertexBound}
$|\bigcup_{S \in {\bf E}_x} S| \leq 2^{x+1}|sm({\bf F})|$.
\end{corollary}

{\bf Proof.}
Observe that $\bigcup_{S \in {\bf E}_x} S=
sm({\bf F}) \cup \bigcup_{i=1}^x \bigcup_{S' \in {\bf E}_i \setminus {\bf E}_{i-1}} hat_i(S')$.
Indeed, by definition, the left set is clearly a superset of the right one, so let $v$
be a vertex of the left set. If $v \in sm({\bf F})$ then the containment in the right
set is clear. Otherwise, let $S^* \in {\bf E}_x$ be a minimal set containing $v$ and let $j>0$ be
the excess of $S^*$. Then, by definition of sets ${\bf E}_i$, 
$S^* \in {\bf E}_j \setminus {\bf E}_{j-1}$. 
From the minimality of $S^*$ subject to the containment of $v$,   
it follows that $v \in hat(S^*)$. Furthermore, by the {\bf SM}  condition,
there is no $S'' \in {\bf E}_j$ such that $S^* \prec S''$. This implies that
$v \in hat_j(S^*)$, confirming the observation. 

It follows from this equality that $|\bigcup_{S \in {\bf E}_x} S| $ is upper-bounded
by $|sm({\bf F})|+\sum_{i=1}^x \sum_{S \in {\bf E}_i \setminus {\bf E}_{i-1}} |hat_i(S)| \leq M_0+\sum_{i=1}^x M_i \leq 
\sum_{i=0}^x M_i$. According to Theorem \ref{MainLevelBound}, the rightmost item of the above inequality is
clearly upperbounded by $2^{x+1}|sm({\bf F})|$, hence the corollary follows. $\blacksquare$

To prove the second statement of Theorem \ref{MainResult}, we need to compute $\bigcup_{S \in {\bf E}_x} S$.
We obtain the required algorithm in four simple steps. First we introduce the notion
of \emph{principal sets} of ${\bf F}$, then we show that the union of principal sets of excess at most
$x$ in fact includes all the vertices of $\bigcup_{S \in {\bf E}_x} S$. Furthermore, we show that 
the number of principal sets can be upper bounded by $2^{x+1}|sm({\bf F})|$. Finally, we show that
subject to {\bf EC} condition, these principal
sets can be computed in time polynomial in their bound and in $n=|V|$. (Recall
that $V$ is the universe of for the sets of ${\bf F}$).

We say that a set $S \in {\bf F}$ is \emph{principal} if $hat(S) \neq \emptyset$.
Denote by ${\bf Pr}_x$ the family of all principal sets of excess at most $x$.
By definition, $\bigcup_{S \in {\bf Pr}_x} \subseteq \bigcup_{S \in {\bf E}_x}$.
For the other direction, let $v \in \bigcup_{S \in {\bf E}_x}$. Then, arguing as
in the proof of Corollary \ref{VertexBound}, we observe the existence of $S^*$
of excess at most $x$ such that $v \in hat(S^*)$. Clearly $S^* \in {\bf Pr}_x$.
Thus we have established the following proposition.

\begin{proposition} \label{TheSame}
$\bigcup_{S \in {\bf Pr}_x} S = \bigcup_{S \in {\bf E}_x} S$
\end{proposition}

\begin{proposition} \label{PrBound}
For each $x \geq 0$, $|{\bf Pr}_x| \leq 2^{x+1}|sm({\bf F})|$. 
\end{proposition}

{\bf Proof.}
By definition, the number of elements of $|{\bf Pr}_x|$
is upper-bounded by the sum of the sizes of their hats, which
in turn, is bounded by the sum of sizes of hats of all elements
of ${\bf E}_x$. Taking into account that for each
$1 \leq i \leq x$ and for each $S \in {\bf E}_i \setminus {\bf E}_{i-1}$,
$hat(S)=hat_i(S)$ (argue as in the proof of Corollary \ref{VertexBound}), 
our upper bound can be represented as
$|sm({\bf F})|+ \sum_{i>1}^x \sum_{S \in {\bf E}_i \setminus {\bf E}_{i-1}} |hat_i(S)|$.
Now, apply the second paragraph of the proof of Corollary \ref{VertexBound}. $\blacksquare$

\begin{theorem} \label{runtime}
${\bf Pr}_x$ can be computed in time $O(n^32^{2x}r^2(r+x)^2)$
where $r=|sm({\bf F})|$.
\end{theorem}

{\bf Proof sketch.}
The algorithm works iteratively. First it computes ${\bf Pr}_0$.
For each $i>0$, it computes ${\bf Pr}_i$ based on ${\bf Pr}_{i-1}$.
Since ${\bf Pr}_0=\{sm({\bf F})\}$, for $i=0$, the result directly follows from
the {\bf EC} condition. Now consider computing of ${\bf Pr}_i$ 
for $i>0$ assuming that ${\bf Pr}_{i-1}$ have been computed. 

The algorithm explores all the elements 
of ${\bf Pr}_{i-1}$  and for each such element $S$ and for each 
$v \in S$, applies the witness computation algorithm 
of the {\bf EC} condition. If the witness
$S'$ of $S$ has been returned, $S'$ joins ${\bf Pr}_i$ if 
$ex(S')=i$, $S'$ has not been already generated and the union
of elements of ${\bf Pr}_i$ preceding $S'$ is not a superset  of $S$.
In the rest of the proof, postponed to the appendix, 
we prove correctness and the runtime of this algorithm. 
$\blacksquare$

\subsection{Back to Important Separators.}
\begin{lemma} \label{conditions}
The family of all important $X-Y$ separators of graph $G$
partially ordered by the $\prec^*$ relation is an IS-family.
\end{lemma}

{\bf Proof sketch.} 
The {\bf SE} condition is established by Lemma 3.3. of \cite{MarxTCS}.
The {\bf SM} condition immediately follows from the definition of an
important separator. For the {\bf SW} condition, let $K$ be an
important $X-Y$ separator and let $v \in K$. Assume that a witness
of $v$ w.r.t. $K$ exists. Replace $NR(G,Y,K)$
by a single vertex $x$ and split $v$ into $n+1$ copies. Let $G^*$
be the resulting graph. We prove that there is a bijection between the
witnesses of $v$ w.r.t. $K$ and smallest important $x-Y$ separators
of $G^*$ and apply to $G^*$ the {\bf SE} condition.
For the {\bf TE} condition, we observe 
(e.g. Proposition 1 of \cite{REX}), that if $K_1 \prec^* K_2$
then $K_1 \setminus K_2 \subseteq NR(G,Y,K_2)$. Thus if  $K_2 \prec^* K_3$,
$K_1 \setminus K_2 \subseteq NR(G,Y,K_2) \subseteq NR(G,Y,K_3)$, the last 
inclusion is obtained by definition of the $\prec^*$ relation. Thus, no
vertex of $K_1 \setminus K_2$ can belong to $K_3$. For the visible set conditions,
we first prove that if $K$ is an important $X-Y$ separator different from
the smallest one then for each $K' \in Pred(K)$, $K^*=Vis(K) \setminus NR(G,Y,K')$ is also
an $X-Y$ separator  such that $K' \preceq^* K^*$.
The {\bf LVS} and {\bf DVS} conditions will immediately follow from this claim
combined with the definition of an important separator. 
The $O(n^3)$ algorithm for computing $sm({\bf F})$, as required by the {\bf EC}
condition follows from Lemma 1 in \cite{REX}. As shown in the proof of the
{\bf SW} condition, computing of a witness is essentially equivalent to
computing of an important separator. Finally the fast testing of $K_1 \prec^* K_2$
is easy to establish by maintaining an important separator in an appropriate
data structure. $\blacksquare$

{\bf Proof of Theorem \ref{MainResult}}
The theorem immediately follows from combination of
Corollary \ref{VertexBound}, Theorem \ref{runtime}, and 
Lemma \ref{conditions}.
$\blacksquare$

\subsection{Lower bounds and possibilities for further improvement}
We start with showing that the obtained upper bound on the number
of vertices involved in important separators of size at most $x$
is quite tight.

\begin{theorem} \label{LowerBound}
For each $x$ and $r$ there is a graph $H$ with two specified terminals $s$
and $t$ such that the size of the smallest $s-t$ separator is $r$ and
the size of the union of all important separators of excess at most $x$
is $2^{x+1}r-r$.
\end{theorem}

{\bf Proof.} Take $r$ complete rooted binary trees of height $x$ with $2^x$ leaves
(of course, replace arcs by undirected edges).
Add two new vertices $s$ and $t$. Connect $s$ to the roots of the trees
and $t$ to all the leaves. This is the resulting graph $H$ for the given
$x$ and $r$. It is not hard to see that any minimal $s-t$ separator of this graph 
is an important one. It only remains to show that each non-terminal vertex
participates in a $s-t$ separator of excess at most $x$. In fact, we can show that
any vertex $v$ whose depth in the respective binary tree is $i$ participates in
a separator of excess $i$. We compute such separator by obtaining a sequence
$S_1, \dots, S_i$ of separators, where $S_i$ is the desired separator.
$S_1$ is just the set of neighbours of $s$. To obtain $S_{j+1}$ from $S_j$, we specify
the unique $u \in S_j$ such that $u$ is the ancestor of $v$ (the uniqueness easily
follows by induction) and replace it by its children. The correctness of this construction
can be easily established by induction on the constructed sequence of separators,
we omit the tedious details. $\blacksquare$

In the previous subsection we introduced the notion of a principal set of an 
IS-family. The corresponding notion of a principal important separator $K$
means that $K \setminus \bigcup_{K' \prec^* K} K' \neq \emptyset$.
Proposition \ref{PrBound} along with Lemma \ref{conditions} implies that the number of 
principal important $X-Y$ separators of excess $x$ is at most $2^{x+1}r$ where $r$ is
the size of the smallest important $X-Y$ separator and the class of graphs considered 
in Theorem \ref{LowerBound} shows that this bound is tight. On the other hand,
the number of principal important separators in this class of graphs is \emph{linear}
in the overall number $n$ of vertices. This leads us to the following question

\begin{openquestion}
Is the number of principal important $X-Y$ separators of the given 
graph $G$ bounded by a polynomial of $|V(G)|$?
\end{openquestion}

First of all observe that this question is reasonable because the
number of principal separators is generally much smaller than the overall
number of important separators. Indeed, in the class of instances considered
in Theorem \ref{LowerBound}, the overall number of important separators is
exponential in $n$ (consider the important separators including leaves 
of the binary trees). 

To see the significance of this open question, suppose that the answer is \emph{yes}.
Then the algorithm claimed in Theorem \ref{MainResult} runs in a polynomial time.
Indeed, its exponential runtime is caused by the fact that the algorithm explores
all pairs of principal important separators, so, replacing the upper bound has an
immediate effect on the runtime. Such poly-time algorithm would mean that it is
possible to test in a polynomial time \emph{whether the given vertex belongs to an important
separator}, which is itself quite an interesting achievement. Moreover, the whole 
preprocessing algorithm for the \textsc{mwc} problem proposed in this paper will have
a polynomial time. This means that the output of this algorithm can be used
for the \emph{further} preprocessing, potentially making easier the unconditional kernelization
of the \textsc{mwc} problem.

\section{Preprocessing of multiway cut}
Let $(G,T)$ be an instance of the \textsc{mwc} problem.
An isolating cut of $t \in T$ is a $t-T \setminus \{t\}$
separator. If 
such separator is important, we call it \emph{important isolating
cut} of $t$.

We start from a proposition that allows us to harness
the machinery of important separators for the 
preprocessing of  the \textsc{mwc} problem. The proposition
is easily established by iterative application the argument
of Lemma 3.6 of \cite{MarxTCS}.

\begin{lemma} \label{OnlyImportant}
Let $(G,T)$ be an instance of the \textsc{mwc} problem.
Then there is a smallest \textsc{mwc} $S$ of $(G,T)$ such that
each $v \in S$ belongs to an important isolating cut of some $t \in T$.
\end{lemma} $\blacksquare$

With Lemma \ref{OnlyImportant} in mind, we can use the algorithm claimed
in Theorem \ref{MainResult} for the preprocessing. In particular,
for each $t \in T$, let $r_t$ be the size of the smallest isolating cut.
Compute the set of all vertices participating in the important isolating
cuts of $t$. Let $V^*$ be the set of all the computed vertices together
with the terminals. Let $G^*$ be the graph obtained from $G[V^*]$ by making
adjacent all non-adjacent $u,v$ such that $G$ has a $u-v$ path with all intermediate 
vertices lying outside $V^*$. It is not hard to infer from Lemma \ref{OnlyImportant}
that the size of the optimal solution of $(G^*,T)$ is the same as of $(G,T)$.
According to Theorem \ref{MainResult}, the number of vertices of 
$G^*$ is at most $|T|(2^{k-r}r+1)$ where $r=min_t \in T r_t$ 
and $1$ is added on the account of terminals. This bound is not good in the sense
that $|T|$ may be not bounded by $k$ at all. Therefore 
\emph{prior} to computing the union of important separators, we reduce the number
of terminals. This is possible due to the following theorem.

\begin{theorem} \label{SqueezeTerminals}
There is a polynomial-time algorithm that transforms the instance
$(G,T,k)$ of the \textsc{mwc} problem into an equivalent instance
$(G',T',k')$ such that $k' \leq k$ and $|T'| \leq 2k'(k'+1)$.
Then runtime of this algorithm is the same as the runtime of 
the fixed-ratio approximation algorithm for the \textsc{mwc} 
problem \cite{Gargapprox} \footnote{This algorithm is based on
solving a linear program.}.
\end{theorem}

{\bf Proof.}
We start from observation that if $u$ is a non-terminal vertex
such that there are $k+2$ terminals connected to $u$ by paths
intersecting only at $u$ then $u$ participates in any 
\textsc{mwc} of $(G,T)$ of size at most $k$. Indeed, removal
of a set of at most $k$ vertices not containing $u$ would leave
at least 2 of these paths undestroyed and hence the corresponding
terminals would be connected. An immediate consequence of this observation
is that if $S$ is a \textsc{mwc} of $G$ and there is $v \in S$ adjacent
to at least $k+2$ components of $G \setminus S$ containing terminals
then this vertex participates in any \textsc{mwc} of $G$ of size at
most $k$.

Having the above in mind, we apply the ratio $2$ approximation
algorithm for the \textsc{mwc} problem proposed in \cite{Gargapprox}. \footnote{In fact,
the approximation ratio of this algorithm is $2-2/|T|$, but ratio $2$ is sufficient
for our purpose.}
If the resulting \textsc{mwc} is of size greater than $2k$, the
algorithm simply returns 'NO'. Otherwise, let $S$ be the resulting
\textsc{mwc}. If $|T|>|S|(k+1)$ then, taking into 
account that each component is adjacent to at least one vertex of $S$,
it follows from the pigeonhole principle that
at least one vertex of $S$ is adjacent to at least $k+2$ components
of $G \setminus S$ containing terminals. 
Remove $v$  and remove isolated components of $G \setminus \{v\}$
(i.e. those that contain at most one terminal), decrease the parameter
by $1$ and recursively apply the same operation to the new data. 
Eventually, one of three possible situations
occur. First, after removal of $k$ or less vertices, the resulting graph
has no terminals. In this case we have just found the desired \textsc{mwc}
of $(G,T)$ in a polynomial time. Second, after removal of $k$ vertices, 
there are still terminals, not separated by the removed vertices. In this
case, again in a polynomial time, we have found that $(G,T)$ has no \textsc{mwc}
of size at most $k$. Finally, it may happen that after removal of some $S' \subseteq S$
of size at most $k$, the number of terminals in the remaining graph is at most $|S \setminus S'|(k-|S'|+1)$.
Then the resulting graph is returned as the output of the preprocessing. $\blacksquare$

Thus, Theorem \ref{SqueezeTerminals} together with Theorem \ref{MainResult} and
Lemma \ref{OnlyImportant} lead to the following result.

\begin{corollary}
There is an algorithm that for an instance $(G,T,k)$ of the \textsc{mwc} problem
finds an equivalent instance of $O(k^2r2^{k-r})$ vertices in time 
$O(A(n)+2^{k-r}n^3r^2k^4)$ where $r$ is the smallest isolating cut and $A(n)$
is the time complexity of the approximation algorithm proposed in \cite{Gargapprox}.
In particular, if $k-r=c*log(k)$ for any fixed $c$ 
then the \textsc{mwc} problem is polynomially kernelizable.
\end{corollary}

The output of the above algorithm is much richer than just another
instance of the \textsc{mwc} problem. Indeed, for each terminal, the 
algorithm in fact computes all principal important isolating cuts.
This leads to the follows interesting question.
\begin{openquestion}
Is there an algorithm that gets the above output as input and, 
in time polynomial in $n$ and the number of the principal isolating cuts,
produces an equivalent instance of the \textsc{mwc} problem of size 
polynomial in $k$?
\end{openquestion}

Observe that if Open Questions 1 and 2 are answered affirmatively then,
together with Proposition \ref{TheSame}, Theorem \ref{runtime}, and Lemma \ref{conditions}, 
they imply an unconditional polynomial kernelization of the \textsc{mwc} problem. Moreover, we believe 
that investigation of Open Question 2 would give a significant insight into 
the structure of the \textsc{mwc} problem. 
Indeed it would reveal whether or not we can 'filter' in a reasonable time 
some principal isolating cuts, which in turn would require proof of some
interesting structural dependencies related to the \textsc{mwc} problem.


\appendix
\section{Proofs omitted from the main body}
{\bf The rest of proof of Theorem \ref{runtime}}
For the correctness, we need to show that the
set of new added elements is precisely ${\bf Pr}_i \setminus {\bf Pr}_{i-1}$.
This is established in the following three paragraphs.

{\bf Every element of $Pr_i \setminus Pr{i-1}$ is collected
     during the gathering stage.}
Let $S \in {\bf Pr}_i \setminus {\bf Pr}_{i-1}$.
Since $hat(S) \neq \emptyset$, by Proposition \ref{PropHat}, $S \neq Vis(S)$.
It follows from the {\bf DVS}  condition that $Vis(S) \nsubseteq S$.
Let $v \in Vis(S) \setminus S$. By Proposition \ref{WitnessAdvanced}, there is
$S^*$ such that $v \in hat(S^*)$ and $S$ is the witness of $v$ w.r.t. $S^*$.
Taking into account the {\bf SM}  condition, we conclude that
$S^* \in {\bf Pr}_{i-1}$, that is $S$ will be generated by the above algorithm.

{\bf An element of $Pr_i \setminus Pr_{i-1}$ will not be filtered out.}
$hat(S)$ consists of elements that are not contained in \emph{any} element
of {\bf F} preceding $S$. The algorithm checks $S$ only against the union of a
\emph{subset} of such elements.

{\bf An element that is not in $Pr_i \setminus Pr_{i-1}$ will be filtered.}
Let $S \in {\bf F}$ be such that $ex(S)=i$ and $hat(S)=\emptyset$.
Let $v \in S$. It is sufficient to show that there is $S^* \in {\bf Pr}_{i-1}$
such that $v \in S^*$ and $S^* \prec S$. Choose $S^*$ to be a minimal element
preceding $S$ and containing $v$. Due to the minimality of $S^*$, $v$
does not belong to any element preceding $S^*$ hence $v \in hat(S^*)$,
i.e. $S^*$ is a principal set. Due to the {\bf SM}  condition,
$ex(S^*) \leq i-1$. It follows that $S^* \in {\bf Pr}_{i-1}$.

Let us compute the runtime. The main cycle goes through all elements of
${\bf Pr}_{i-1}$, the number of such elements is at most
$2^{x+1}r$ according to Proposition \ref{PrBound}.
Let us compute the time spent per element.
Since each element of ${\bf Pr}_{i-1}$ is of excess at most $x$,
i.e. of size at most $r+x$, the algorithm explores at most $r+x$
vertices and for each vertex either computes the respective witness
or concludes its absence. It follows from the {\bf EC} condition 
that the overall time spent for
computation of witnesses \emph{per} element of ${\bf Pr}_{i-1}$ is
$O(n^3(r+x))$. Let us compute time spent \emph{per} witness.
Denote the considered witness by $S_1$. Then $S_1$ is compared against
all elements of ${\bf Pr}_{i-1}$ where the number of such elements is
at most $2^{x+1}r$ as noticed above. For each $S_2 \in {\bf Pr}_{i-1}$,
it is checked whether $S_2 \prec S_1$ which can be done in $O(r+x)$
according to the {\bf EC} condition. If the test returns a positive
answer then $S_2$ is added to the union of elements preceding $S_1$ which,
using appropriate data structures \footnote{the union can be stored as a binary
vector of size $n$ indexed by the elements of the universe and adding a set to the
union just means ticking the respective entries $|S_1|$ times}
can be done in $O(|S_2|)$, i.e. again in $O(r+x)$. Thus the total runtime
of this operation is $O(2^xr(r+x))$. After finishing the comparison against
${\bf Pr}_{i-1}$, the algorithm checks whether or not all the elements are
in the resulting union of predecessors. This can be done in $O(|S_1|)$ i.e. in
$O(r+x)$, clearly this runtime can be ignored in the light of the already
spent $O(2^xr(r+x))$. Multiplying the number of considered witnesses by the runtime
spent per witness, the desired runtime of $O(n^32^xr^2(r+x)^2)$ follows. $\blacksquare$

{\bf Proof of Lemma \ref{conditions}}
We show that the set of important separators partially ordered by the $\prec^*$
meets all the conditions of the IS-family.

{\bf SE Condition} See Lemma 3.3 of \cite{MarxTCS}.\\
{\bf SM Condition} Immediately follows from
the definition of an important separator. \\
{\bf SW Condition} Let $K$ be an important separator.
Let $G^*$ be the graph obtained from $G$ by contraction of all the
vertices of $NR(G,Y,K) \setminus X$. In other words, to obtain
$G^*$ from $G$, remove all vertices of $NR(G,Y,K)$ and add
an edge between each vertex $u \in X$ and $v \in K$ such that
there is a $u-v$ path all intermediate vertices of which belong to
$NR(G,Y,K)$. It is not hard to see that the definition of an important
separator implies that $K$ is the smallest $X-Y$ separator of $G^*$.
Let $v \in K$. Assume that $v$ is not adjacent to $Y$ in $G$ (otherwise
there is no witness of $v$ w.r.t. $K$). Let $G''$ be a graph obtained from $G^*$
by splitting $v$ into $n+1$ copies. It is not hard to observe that
the set of important $X-Y$ separators of $G''$ is the set of
important $X-Y$ separators of $G$ that do not contain $v$, moreover
the partial order relation is preserved. Then a witness of $K$ w.r.t. $v$
in $G$ is a smallest important $X-Y$ separator of $G''$. By the {\bf SE}
condition, this separator is unique.
\\
{\bf TE condition.}
Observe (e.g. Proposition 1 of \cite{REX}), that if $K_1 \prec^* K_2$
then $K_1 \setminus K_2 \subseteq NR(G,Y,K_2)$. Thus if  $K_2 \prec^* K_3$,
$K_1 \setminus K_2 \subseteq NR(G,Y,K_2) \subseteq NR(G,Y,K_3)$, the last
inclusion is obtained by definition of the $\prec^*$ relation. Thus, no
vertex of $K_1 \setminus K_2$ can belong to $K_3$.

In order to establish the visible set conditions, we  prove an intermediate claim.
\begin{claim}
Let $K$ be an important $X-Y$ separator, which is not the smallest
one and let $K' \in Pred(K)$. Then $K^*=Vis(K) \setminus NR(G,Y,K')$ is
a $X-Y$ separator such that $K' \preceq^* K^*$.
\end{claim}

{\bf Proof.}
Let $v \in NR(G,Y,K')$ and let $p$ be a $v-Y$ path of $G$.
Let $u$ be the last vertex of $p$ that belongs to $\bigcup_{K'' \in Pred(K)} K''$.
Clearly, $u$ is not covered by any element of $Pred(K)$ because
otherwise it would not be the last vertex of $p$ that belongs to an element of $Pred(K)$. 
Hence by definition $u \in Vis(K)$. Clearly, $u$ cannot belong to $NR(G,Y,K')$ because
otherwise it will be followed in $p$ by an element of $K'$. Consequently, 
$u \in Vis(K) \setminus NR(G,Y,K')$, confirming the claim. $\square$

{\bf LVS condition.} According to the above claim $|K'| \leq |K^*|$
because otherwise we get a contradiction to being $K'$ an important separator.
Taking into account that $K^* \subseteq Vis(K)$, the condition follows.

{\bf DVS condition.} $K^* \subseteq Vis(K)$, hence the latter is an $X-Y$ 
separator. Therefore, if $Vis(K) \subset K$ then $K$ is not a minimal $X-Y$
separator in contradiction to its importance.

{\bf EC condition}
The $O(n^3)$ algorithm for computing a smallest important separator
follows from Lemma 1 in \cite{REX}.
This immediately implies existence of such algorithm for the witness
computation. Indeed, the single witness
condition proof of Lemma \ref{conditions} shows that witness
computation can be reduced to computing the smallest important
separator and such the reduction can be clearly performed in $O(n^3)$.
Finally, observe that it is possible to maintain an important separator
$K$ in a way that for each vertex $v$ testing whether $v \in NR(G,Y,K)$
can be performed in $O(1)$: associate $K$ with a binary vector of size
$n$ indexed by $V(G)$ where $1$-s correspond to the elements of $NR(G,Y,K)$.
In the light of Proposition 1 in \cite{REX}, this immediately implies
that $K_1 \prec^* K_2$ can be tested in $O(K_1)$. $\blacksquare$

{\bf Proof of Lemma \ref{OnlyImportant}} Let $S_1$ be an arbitrary smallest \textsc{mwc} of
$(G,T)$. If all vertices of $S_1$ belong to important isolating cuts,
we are done. Otherwise, let $v \in S_1$ be a vertex that does not belong
to any smallest isolating cut. Due to the minimality of $S_1$,
there is $t \in T$ such that $v$ belongs to a minimal isolating cut
$S' \subseteq S_1$ of $t$. It follows that there is an important
isolating cut $S'' \succ^* S'$ of $t$ such that $|S''| \leq |S'|$.
The proof of Lemma 3.6. of \cite{MarxTCS} shows that
$S_2=(S_1 \setminus S') \cup S''$ is a \textsc{mwc} of $(G,T)$ of
size not exceeding $S_1$. In other words $S_2$ is an optimal solution
of $(G,T)$  and the number of vertices of $S_2$ not involved in
any important isolating cuts is smaller than that of $S_1$.
Applying such modification iteratively, we eventually obtain a smallest
multiway cut without such 'undesired' vertices. $\blacksquare$
\end{document}